\documentclass[manuscript,nonacm]{acmart}

\usepackage[utf8]{inputenc}
\usepackage{comment}
\usepackage{listings}
\usepackage{xcolor}

\acmBooktitle{arXiv}

\title{QUANTAS: Quantitative User-friendly Adaptable Networked Things Abstract Simulator }
\author{Joseph~Oglio, Kendric~Hood, Mikhail~Nesterenko}
\email{joglio@kent.edu, khood5@kent.edu, mikhail@cs.kent.edu}
\affiliation{\institution{Kent State University} \country{Kent, USA}}
\author{S\'{e}bastien~Tixeuil}
\email{Sebastien.Tixeuil@lip6.fr}
\affiliation{\institution{Sorbonne Universit\'{e}, CNRS, LIP6} \country{FR-75005, France}}
\date{\today}
\begin{document}

\begin{abstract} We present QUANTAS: a simulator that enables quantitative performance analysis of distributed algorithms. It has a number of attractive features. QUANTAS is an abstract simulator, therefore, the obtained results are not affected by the specifics of a particular network or operating system architecture. QUANTAS allows distributed algorithms researchers to quickly investigate a potential solution  and collect data about its performance. QUANTAS programming is relatively straightforward and is accessible to theoretical researchers. 
To demonstrate QUANTAS capabilities, we implement and compare the behavior of two representative examples from four major classes of distributed algorithms: blockchains, distributed hash tables, consensus, and reliable data link message transmission.
\end{abstract}
\maketitle

\section{Introduction}
Theoretical work in distributed algorithms often involves establishing a possibility of solution existence, proving an algorithm correct or determining its message or computation complexity. If the new algorithm improves on the existing ones, this improvement is quantified in terms of these complexity metrics.
These approaches may be lacking as they do not provide sufficient insight into the realistic behavior of the algorithm. Indeed, hidden constants and system parameters, such as message delay or relative computation power, influence the performance of most algorithms.

Alternatively, the algorithm is implemented in 
a real distributed architecture such as a computer cluster, a collection of virtual machines, a cloud computing service \cite{azure}, or using a general purpose network simulator such as ns-3~\cite{ns3} or OMNET++~\cite{omnetpp}. Although such efforts demonstrate practical algorithm implementation and enable its immediate application, the obtained results make it difficult to separate the operation of the algorithm from the influence of the particular network and operating system used in performance evaluation. For example, it is unclear how the interaction between virtual machines and network switches at the server farm affects the results obtained in real network performance evaluation or whether the selection of a particular physical layer network protocol made a difference in a network simulator.

Another obstacle for these approaches is difficulty of performing large-scale performance evaluation. Large scale physical systems are expensive to procure for experimentation. Moreover, in a physical system, there is difficulty instrumenting and then ascertaining conditions of interest for experimenter such as specific network delay. Network simulators, due to extensive simulation detail, also have limited scalability.

To demonstrate the behavior of a distributed algorithm and compare it with the alternatives, abstract simulation may be used. Abstract simulation closely follows the communication and computation model used in distributed algorithms research. The algorithm is represented as a collection of processes and communication channels or shared variables. The computation is modeled as series of simulation rounds where processes perform concurrent processing and exchange messages. Such modelling of algorithms make abstract simulation attractive to distributed algorithms researchers. 

However, we believe there is a lack of general purpose tools for such abstract simulation. Most papers use 
ad hoc one-off implementations built for one paper~\cite{hood2021partitionable,fastGeometric,adamek2015evaluating,adamek17jpdc} or, at best, a domain specific abstract simulation that is reused for a limited number of papers~\cite{peersim,adamek17srds,adamek18icdcn}. 
This duplicates effort and makes it difficult to verify obtained results, compare them across several publications, or make further improvements. The simulation code and obtained data are seldom made publicly available which further exacerbates the problem. 

The existing general purpose abstract simulators, that we are aware of, tend to be used for education rather than research. 
However, we think that the focus of an educational simulator differs from that of a research simulator. Indeed, the major concern of an educational simulator is to give novices an exposure to the distributed algorithm operation, and a visual representation of the algorithm as it executes~\cite{lydian,jbotsim,sinalgo}. Therefore, simplicity and ease of use are of primary importance. While simplicity certainly does not harm a distributed algorithm simulator, other important characteristics such as scalability, simulation speed, versatility, and ability to obtain quantitative measures for metrics of interest come to the fore.
The closest simulation framework we could find is Neko~\cite{urban2001neko}, however the project seems no longer maintained and its source code no longer available.

In this paper, we present QUANTAS: an abstract simulator specifically designed for distributed algorithms research.  Our primary research area is distributed algorithms. The development of QUANTAS arose out of our own need to do performance evaluation. We, therefore, built QUANTAS to satisfy the needs of researchers similar to our own. We used QUANTAS prototype in several studies~\cite{blockguard,hood2021partitionable}. The QUANTAS software is publicly available~\cite{QUANTASgithub} for other researchers to download and use. 

\section{Simulator Design Principles, Architecture and Setup}

\textbf{Design principles.} 
\begin{itemize}
\item The foremost principle is simplicity, ease of use, and ability to obtain quantitative results quickly. That is, a newcomer should be able to implement algorithms, and get simulation data in a relatively straightforward manner. Quantas interfaces remain basic for ease of integration with analysis or input generation tools. The simulator core is coded in C++. No further compilers, libraries, or specialized languages for distributed algorithm specification are used. In our experience, the benefits of such constructs are limited: what is gained it simplicity is lost in flexibility and speed.
Parameter customization and experiment set up is done through simple text-based configuration files. 

\item As a starting point and demonstration of the simulator capabilities, we provide a set of representative examples. We think these examples will be used with minor modifications by a majority of QUANTAS users for their own research experimentation. 

\item Once the basic behavior of a distributed algorithm is ascertained, the researchers usually want to observe its behavior at scale: large system size, extensive simulated time or resource usage. To support this, we implemented QUANTAS in C++ with minimum overhead.
C++ threading is used to implement concurrent simulation of multiple procsses.
Potentially, the simulated network size is limited by the host computer processor and memory resources.

\item The major simulation goal is to obtain data for analysis and presentation.  QUANTAS is configured to output simulation data in JSON format for ease of further processing. One can then use various available interactive or automated tools to analyze and plot the data.

\item QUANTAS combines all these features in a relatively compact, modular codebase which is easily extensible and modifiable. QUANTAS contains approximately $4,000$ lines of C++ code.

\end{itemize}

\begin{figure}
   \centering
    \includegraphics[trim=350 0 100 0 left bottom right top, clip, height=10cm]{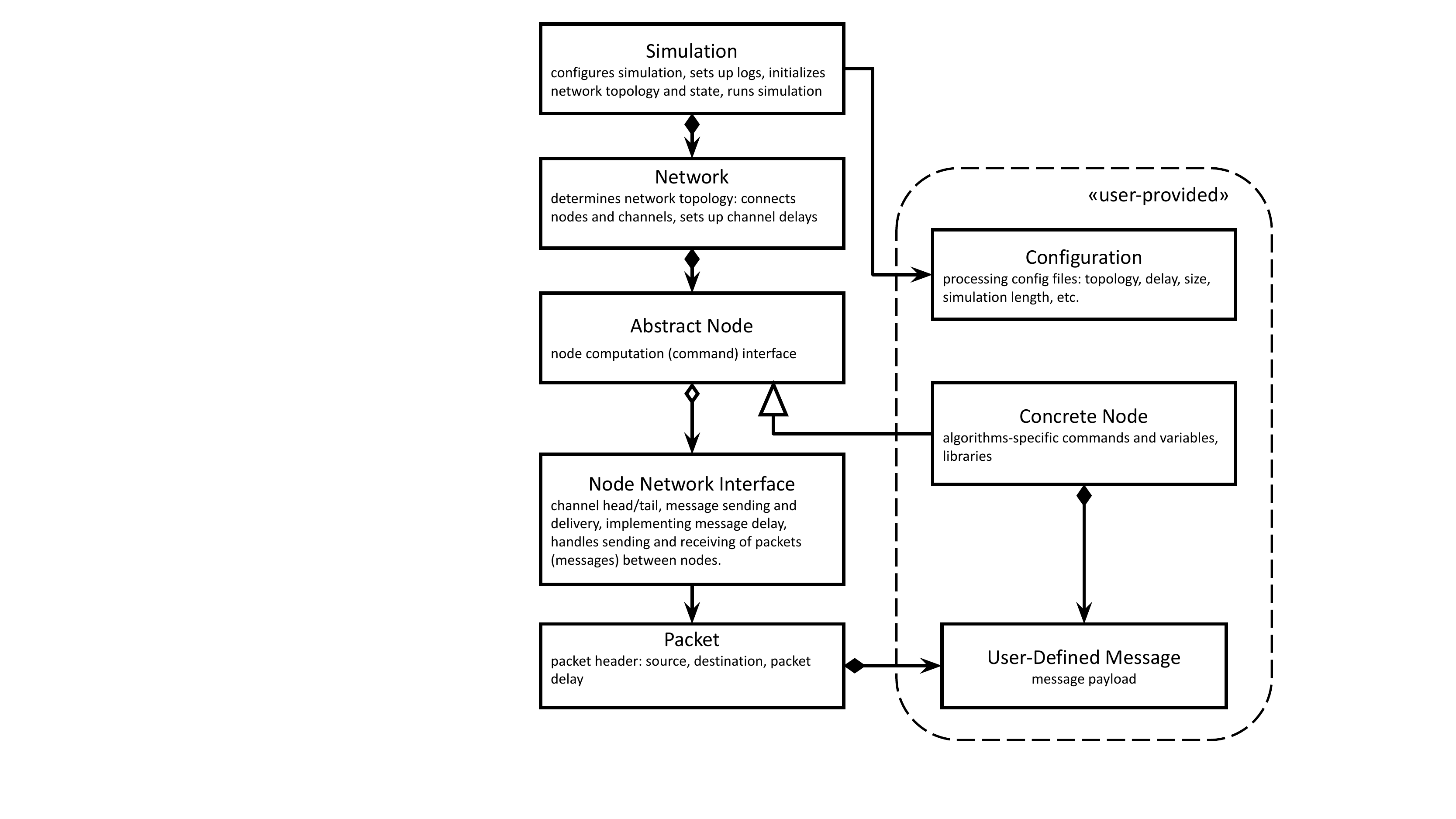}
     \vspace{-6mm}
    \caption{QUANTAS architecture.}
    \label{figArch}
\end{figure}

\noindent
\textbf{Terms and operation.}
A simulated \emph{distributed algorithm} operates on a list of nodes connected via unicast channels. Each channel connects a single sender and a single receiver.  Every node has a unique identifier. 
Each \emph{computation} of the distributed algorithm is a sequence of receive-compute-send \emph{rounds}. Each round has three \emph{phases}: receive messages, perform local computation, and send newly formed messages. A computation \emph{length} is its number of rounds.
A message takes at least one round to pass between nodes that are directly connected through a channel. A message can be delayed. Delay length is configured. The delay is also configured to be either deterministic, uniformly random, or following a Poisson distribution. Communication channels are FIFO by default. Other message propagation delay disciplines may be added by the user.
A transmitted message may be configured to be lost with a certain probability. A message may be sent to an individual process or broadcast to the entire network.
A single \emph{run} of the QUANTAS simulator executes several algorithm computations with the same parameters. This allows QUANTAS to execute multiple individual experiments for a single data point.

\begin{figure}
    \centering
    \begin{lstlisting}
struct HelloMessage {
	std::string messageText;
};
...
void performComputation() {	
    HelloMessage msg;
    msg.messageText = "Hello From " + std::to_string(id()); // send "hello" to all processes
    broadcast(msg); 

    // service all received messages
    while (!inStreamEmpty()) {
        Packet<HelloMessage> newMsg = popInStream();

        // logger is a Singleton, "Greetings" is a tag
        // getRound() returns simulated computation round
        LogWriter::instance()->
            data["Greetings"].push_back(newMsg.getMessage().messageText + "at round: " + getRound())
    }
}
\end{lstlisting}
    \caption{Example QUANTAS code for a local computation phase: each process broadcasts a single message and receives messages from its neighbors.}
    \label{figCode}
\end{figure}

\ \\
\textbf{Architecture.} 
QUANTAS architecture is shown in Figure~\ref{figArch}. The components represent the larger C++ templates and classes. 
The components are in two categories: user-provided and the simulator proper. The user-provided components encode the algorithm to be simulated. The simulator proper components carry out the simulation. The run-time operation of the simulator is controlled by configuration files.

The  \emph{Simulation Component} configures and initializes the simulation run. It then carries out the receive-compute-send computation rounds of individual computations of the run. The Simulation Component uses the \emph{Configuration Component} for processing user-supplied configuration file containing network topology and size, parameters of the run, message delay discipline and parameters, computation length, etc. The network topology is specified as adjacency list and can be generated by hand or by a separate tool. 
This component maintains a thread pool to concurrently execute rounds of multiple simulated processes. 

The \emph{Network Component} configures distributed algorithm topology, sets up communication channels and executes receive- compute- and send- phases of the round. The \emph{Abstract Node Component} is a C++ abstract class that lists the interfaces to be implemented by a user-provided \emph{Concrete Node} component. The main part of this interface is the code to be executed in local computation phase of the round.  

The \emph{Node Network Interface Component} executes receive and send phases of the round. In the receive phase, The \emph{Node Network Interface Component} examines all the channels, and determines if any of the messages currently in transit are ready to be received. The ready-to-receive messages are made available for the computation phase. If the computation phase generates messages to be transmitted, the \emph{Node Network Interface Component} collects them and puts them in the appropriate destination channels.

A message is enclosed in a packet. The packet contains the source, destination, and the delay for this particular message. The \emph{Packet Component} provides this header and the \emph{Node Network Interface Component} uses this header for message routing. The actual message format and its payload are provided by the \emph{User-Defined Message Component}. 

Let us discuss QUANTAS data output capabilities. QUANTAS provides a global logging facility, so that each component may output to the log file. All simulator components may output data about their particular operation. For example, the \emph{Node Network Interface Component} may output sender and receiver identifiers for each individual message. The user-provided components may output arbitrary data, which enables user-specific metrics to be easily implemented. User-provided components have access to the computation round number maintained by the simulator. This round number can be included in the output for analysis. To simplify later processing, logger allows attach an arbitrary tag to output lines. 
QUANTAS example code is shown in Figure~\ref{figCode}.

\section{Examples}

In this section, we demonstrate how QUANTAS may be adapted to fit diverse experimentation needs of distributed algorithms researchers. We chose four domains and a pair of previously published well-known algorithms in each. We then implemented the algorithms in QUANTAS and compared their performance. While the results themselves are not surprising, they demonstrate how QUANTAS may be used for performance evaluation of a variety of algorithms.
\begin{figure}
   \centering
    \includegraphics[width=8cm]{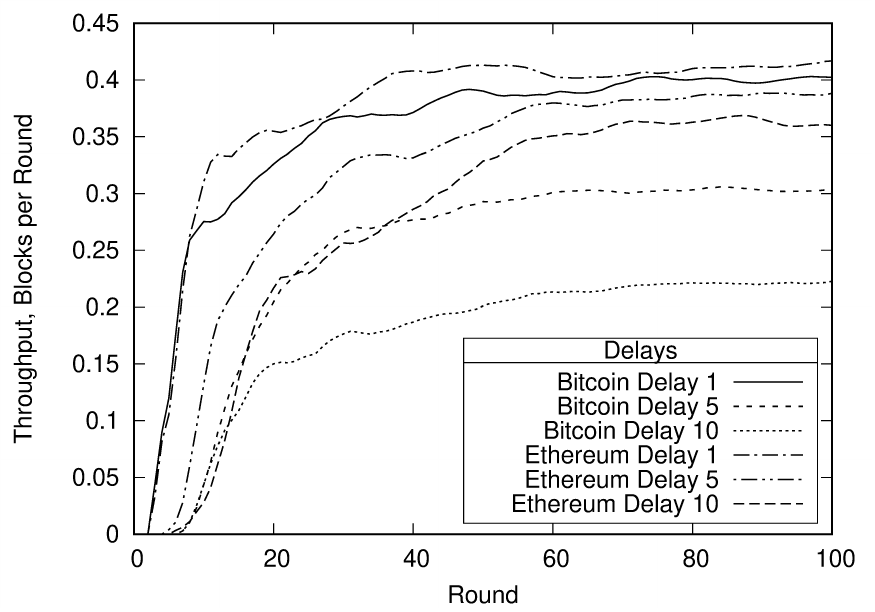}
    \caption{Blockchain throughput during a computation.}
    \label{figBlockchain}
\end{figure}

\ \\
\noindent\textbf{Blockchains.} Blockchain is a secure distributed ledger maintained by a network of peers that compete to add blocks of transactions to the tail of the chain. We simulate simplified versions of the two most widely used blockchain algorithms: Bitcoin~\cite{bitcoin} and Ethereum~\cite{ethereum}.  The peer-to-peer system has 20 peers. Each peer maintains its own copy of the blockchain.  A transaction is submitted to a random peer with probability $5\%$ per peer per round, i.e. on average, it is one transaction per round. The peer receiving the transaction broadcasts it to the rest of the network. In Bitcoin, each peer mines one of the received transactions attempting to link it to the longest chain. The mining probability for each peer is $2.5\%$. In Ethereum, each block links to all previously unlinked blocks. The single computation was executed for $100$ rounds. Each simulator run had $10$ experiments.

The results are shown in Figure~\ref{figBlockchain}. We estimate the number of confirmed blocks by considering the longest chain for each peer and determining the shortest among those. For each blockchain algorithm, we executed a computation for $100$ rounds and calculated the average number of blocks per round. Figure~\ref{figBlockchain} shows a moving average of this value with a window of $5$ rounds. The results shown for message delays $1$ through $10$. The results indicate that our implementation of Ethereum has better throughput than Bitcoin since Ethereum, unlike Bitcoin, may confirm multiple competing blocks concurrently. 

\begin{figure}
\begin{tabular}{c c}
\begin{minipage}[t]{0.49\textwidth}
\includegraphics[width=\textwidth]{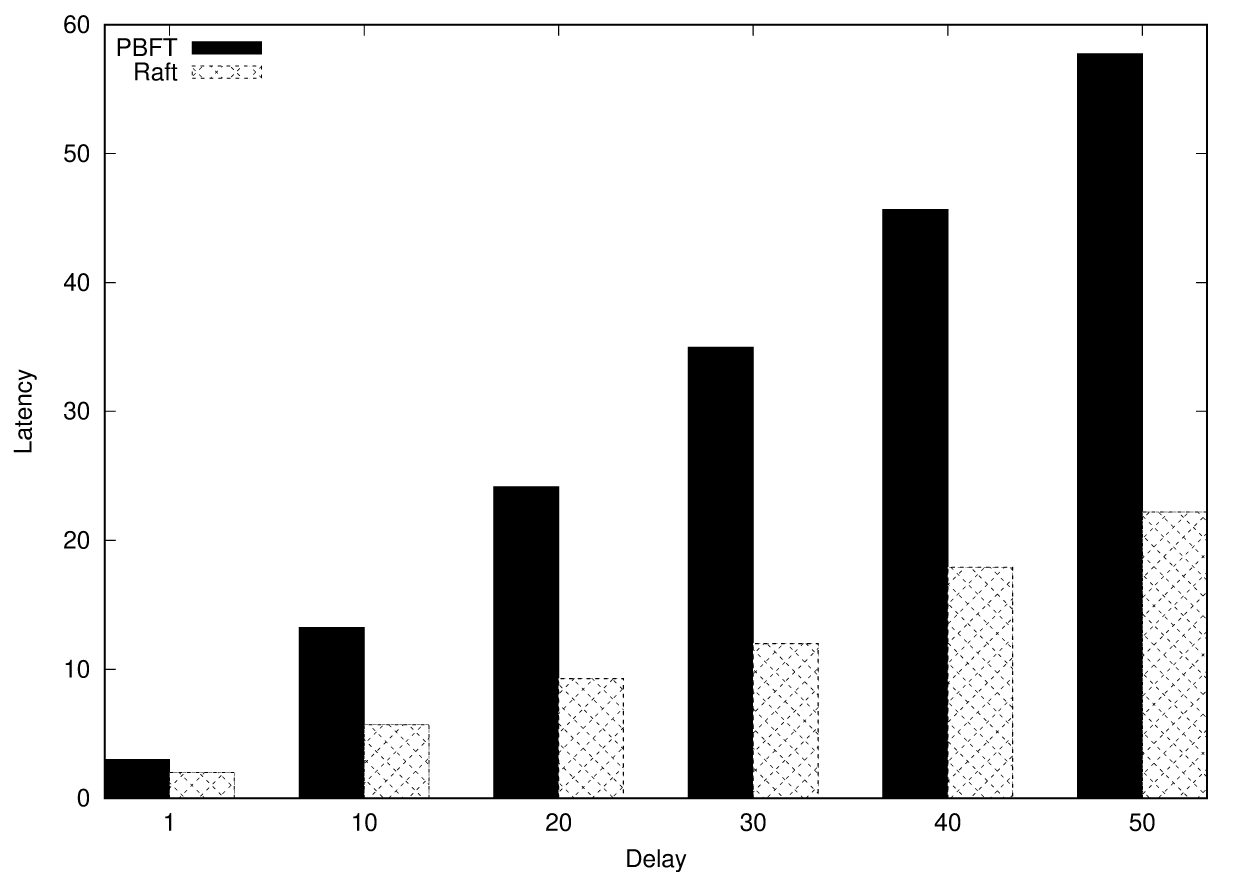}
\caption{Consensus. Latency of achieving a single decision vs message delay.}
\label{figConsensus}
\end{minipage}
&
\begin{minipage}[t]{0.49\textwidth}
\includegraphics[width=\textwidth]{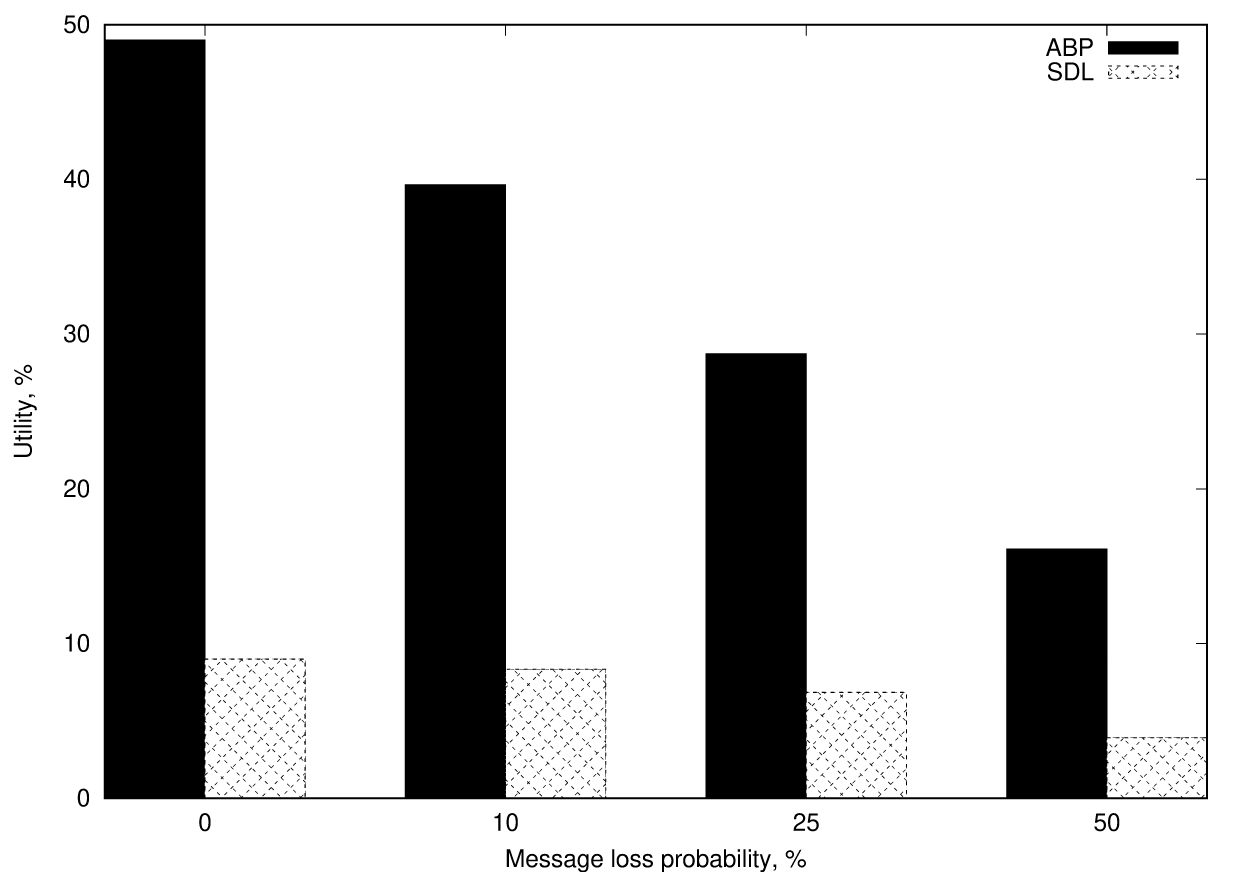}
\caption{Reliable Data Link. Message utility depending on message loss.}
\label{figRDL}
\end{minipage}
\end{tabular}
\end{figure}

\ \\
\noindent\textbf{Robust Consensus.} In robust consensus, a network of nodes attempts to agree on a single input value. We simulated two resilient consensus algorithms: PBFT~\cite{PBFT} and Raft~\cite{raft}. Both algorithms process a sequence of consensus requests. PBFT is resilient to Byzantine faults~\cite{lamport2019byzantine}. In our implementation of PBFT, there is a fixed leader process $l$. The leader $l$ has a sequence of values to commit. For the confirmation, $l$ consults the rest of the processes. For each individual value, $l$ executes PBFT. Specifically, $l$ broadcasts \emph{pre-prepare} message to all processes with the value to be committed. Once a processes receive sufficiently many \emph{prepare} messages with the same value, it commits the value and sends \emph{commit} message informing everyone of this. After receiving  sufficiently many \emph{commit}s, $l$ considers this PBFT instance terminated and moves on to committing the next value. The leader change is not implemented.

Raft~\cite{raft} is resilient to crashes and churn but not to Byzantine faults. 
In Raft, the leader $l$ broadcasts requests to all nodes in the system and waits for majority of responses. After receiving this majority, $l$ moves to the next value to be committed. 

The commitment \emph{latency} is the number of rounds it takes the algorithm from the moment the initial message is transmitted by the leader until the last required commit message is received by the leader.  We used $20$ processes, we executed a computation for $1,000$ rounds. Each simulator run had $10$ computations. We average commitment latency across the run. We varied message delay and recorded the latency of RAFT and PBFT. The results are shown in  Figure~\ref{figConsensus}. RAFT has significantly lower commitment latency than PBFT as there are fewer rounds of message exchanges. This speed is obtained at the expense of resiliency to Byzantine faults. 

\ \\
\noindent\textbf{Reliable Data Link.} In a data link algorithm, the sender process attempts to transmit data to the receiver process despite message loss in the communication channel. A self-stabilizing algorithm~\cite{dijkstra1974self} is resilient to global state corruption. We implemented two self-stabilizing data link algorithms: alternating-bit protocol (ABP)~\cite{howell2002finite} and stabilizing-data link algorithm (SDL)~\cite{dolev2011stabilizing}. ABP requires FIFO channels. SDL operates correctly even in non-FIFO channels. In ABP, the sender transmits a single data message and waits for acknowledgement from the receiver. If either the data message or the acknowledgement is lost, the sender times out and retransmits the message.
In SDL, to enable the receiver to get messages in correct order in a non-FIFO channel, the sender transmits the same message multiple times. The number of transmissions is determined by maximum channel size. 

To compare the two algorithms, we used channels of size one. For SDL, this channel size means that the sender sends the same message $5$ times. 
In our simulation,  we used $2$ processes: the sender and the receiver.  We executed the computation for $100$ rounds. Each simulator run was $10$ computations. We computed \emph{message utility} --- the ratio of successfully received message over transmitted. We varied message loss rate and recorded the utility of the two algorithms. 

The results are shown in Figure~\ref{figRDL}. In our simulation, as message loss increased, both algorithms had to submit more messages to get the data across. This means that the utility decreased for both algorithms. However, SDL effectively submitted about five times as many messages as ABP. This is the overhead needed by the SDL to enforce sequential message delivery in a non-FIFO channel.  

\begin{figure}
\begin{tabular}{c c}
\begin{minipage}[t]{0.49\textwidth}
\includegraphics[width=\textwidth]{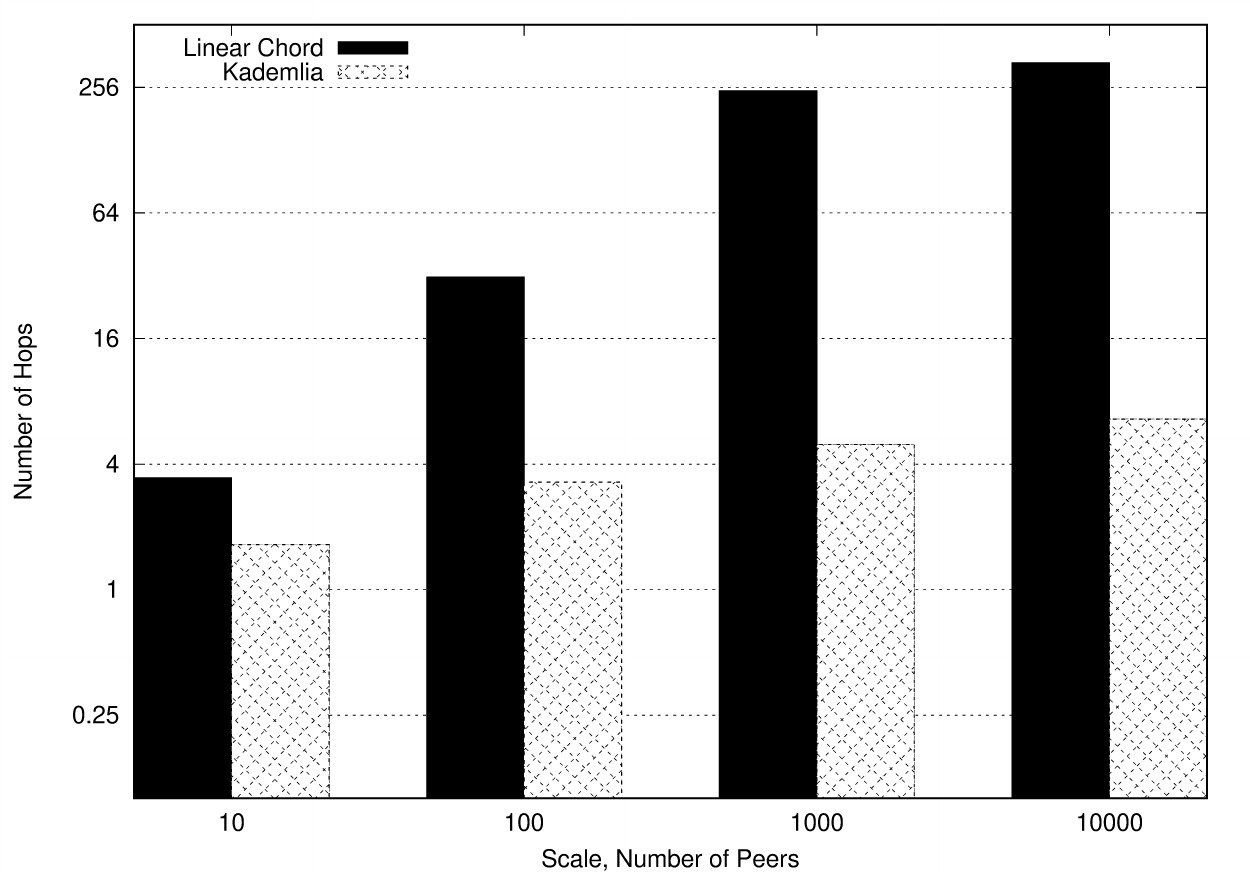}
\caption{Distributed Hash Table query speed.}
\label{figDHT}
\end{minipage}
&
\begin{minipage}[t]{0.49\textwidth}
\includegraphics[width=\textwidth]{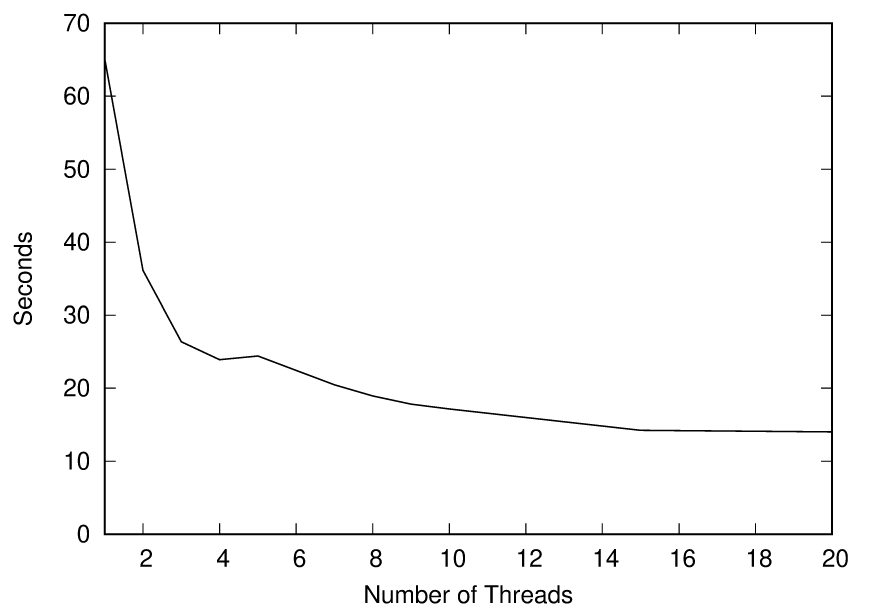}
\caption{Distributed Hash Table simulation speedup.}
\label{figSpeedup}
\end{minipage}
\end{tabular}
\end{figure}

\ \\
\noindent\textbf{Distributed Hash Tables.} In a distributed hash table (DHT), a peer-to-peer system provides query service for key to data items spread throughout the network. The algorithm is optimized to minimize the number of lookups per query. Some of the most widely used DHTs are Chord~\cite{chord} and Kademlia~\cite{kademlia}.

In our Chord implementation, the peer identifiers form a ring. Shortcut links are not implemented. A query for an identifier chosen uniformly at random is generated by another random node. The query is routed to the destination node in the shortest direction. 

In Kademlia, on top of this basic Chord implementation, we also build shortcut links as follows. Peer identifiers are treated as bit fields. A prefix peer group for a particular peer $p$ is a set of peers whose identifiers share a prefix of particular length $l$ with $p$ and differ from $p$ at length $l+1$. For example, if the prefix is one bit less than the complete id length, then, there is a single member in this peer group. A peer group for a prefix that is two bits shorter than id length, contains two members. A peer group three bit shorter than id length contains 4 members and so on.  For each group, a peer selects a random member and creates a shortcut link to it. The query routing is as follows. The peer selects a member with the closest prefix to the destination and routes the query there. 

The results are shown in Figure~\ref{figDHT}. The results indicate that our implementation of Kademlia outperforms Chord because Kademlia query routes are logarithmic with respect to the network size.

To test QUANTAS parallel performance, we varied the number of threads in the simulator thread pool and measured the runtime of the simulation. We simulated $500$ peers. Each computation was $100$ rounds. We run $10$ computations per data point. 
The simulator used approximately $40$ GB of RAM and ran on a virtual machine with a host machine having 2 Intel Xeon Gold 6132 CPUs running at 2.60 GHz.
The virtual machine had 12 cores. The results are shown in Figure~\ref{figSpeedup}.
The results indicate that the simulation speed increases as more threads are added to the simulator thread pool. This speed increase ends as all available host processor cores are used for the concurrent simulation. 

\section{Future Work}

We anticipate further enhancements of QUANTAS capabilities. Besides already implemented message loss, we would like to add more sophisticated fault injection. In particular, we plan to add support for self-stabilizing algorithms evaluation. Even though a self-stabilizing algorithm is proven to recover from an arbitrary global state, evaluating the algorithm's performance starting from a state generated uniformly at random is not realistic as not all such states are equally likely to appear. A more sophisticated approach was developed by Adamek et al~\cite{adamek12sss}: an achievable state of a self-stabilizing algorithm is selectively perturbed. We would like to implement this kind of fault-injection in QUANTAS. Adding crash faults would be a simple and useful addition. A more challenging task is adding Byzantine faults since Byzantine processes are expected to behave so as to inflict the most damage to the algorithm. Hence, despite extensive studies of Byzantine fault tolerance, few of them have performance evaluation. We believe adding Byzantine fault injection~\cite{asim10mswim} would be helpful to the research community. 

We would like to add random topology generation that QUANTAS so that the processes and channels are configured randomly according to the topology graph parameters provided in the configuration file, for example a random graph with the specified node number and edge probability. 
Another feature we find useful is facilitation of application level separation. This would allow the simulator to evaluate levels of multi-level algorithms separately, for example, evaluate the same consensus algorithm over different broadcast algorithms. 

Another important issue is scalability increase, be it related to the number of simulated nodes~\cite{asim10wowmom} or the number of simulations per data point to obtain quantitative results. 
To improve the performance of QUANTAS at scale, we plan to pursue multithreading inside the simulator more aggressively. For example, by using parallel algorithms in the Standard Template Library of C++. In the future, we would like to explore distributed multi-computer simulation.

\ \\
In this paper, we presented QUANTAS, a general abstract simulator dedicated to distributed algorithms quantitative evaluation. While we provided a number of case studies, we welcome contributions from the Distributed Computing community, to build a library of ready-to-use templates for most algorithmic paradigms, that enables fair comparison with previous work when designing new solutions. We believe that QUANTAS fulfils the need for an abstract simulator among researchers of distributed algorithms and we hope it proves to be useful and convenient.

\begin{acks}
We are thankful to Shishir Rai of Kent State University for providing code for an example as the simulator was being developed.
\end{acks}

\bibliographystyle{plain}
\bibliography{quantas}

\end{document}